\documentclass[epj,nopacs]{svjour}
\usepackage{amsmath}
\usepackage{amssymb}
\usepackage{epsfig}
\usepackage{color}

\newcommand{\beq}  {\begin{equation}}
\newcommand{\eeq}  {\end{equation}}

\newcommand{\etal}{\textit{et~al.}}
\newcommand{\JOURNAL}[4]{#1 {\bf #2}, #3 (#4)}

\newcommand{\EPJC}[3]{\JOURNAL{Eur. Phys. J. C}{#1}{#2}{#3}}
\newcommand{\EPJA}[3]{\JOURNAL{Eur. Phys. J. A}{#1}{#2}{#3}}
\newcommand{\PRL}[3]{\JOURNAL{Phys. Rev. Lett.}{#1}{#2}{#3}}
\newcommand{\PRC}[3]{\JOURNAL{Phys. Rev. C}{#1}{#2}{#3}}
\newcommand{\PRD}[3]{\JOURNAL{Phys. Rev. D}{#1}{#2}{#3}}
\newcommand{\PLB}[3]{\JOURNAL{Phys. Lett. B}{#1}{#2}{#3}}
\newcommand{\NIMA}[3]{\JOURNAL{Nucl. Instrum. Methods A}{#1}{#2}{#3}}

\newcommand{\NPB}[3]{\JOURNAL{Nucl. Phys. B}{#1}{#2}{#3}}
\newcommand{\ZPC}[3]{\JOURNAL{Zeit. Phys. C}{#1}{#2}{#3}}
\newcommand{\PR}[3]{\JOURNAL{Phys. Rept.}{#1}{#2}{#3}}
\newcommand{\PRv}[3]{\JOURNAL{Phys. Rev.}{#1}{#2}{#3}}

\newcommand{\CPC}[3]{\JOURNAL{Comput. Phys. Commun.}{#1}{#2}{#3}}

\newcommand{\SJNP}[3]{\JOURNAL{Sov. J. Nucl. Phys.}{#1}{#2}{#3}}
\newcommand{\Nat}[3]{\JOURNAL{Nature}{#1}{#2}{#3}}
\newcommand{\desy}{DESY}

\usepackage{ifpdf}
\ifpdf
\DeclareGraphicsExtensions{.pdf, .jpg, .tif}
\usepackage[%
  pdftitle={Bose--Einstein correlations in hadron-pairs from lepto-production on nuclei ranging from hydrogen to xenon},%
  pdfauthor={The HERMES Collaboration},%
  pdfsubject={HERMES BEC},%
  pdfstartview=FitH,%
  bookmarks=true,%
  bookmarksopen=true,%
  breaklinks=true,%
  colorlinks=true,%
  linkcolor=blue,anchorcolor=blue,%
  citecolor=blue,filecolor=blue,%
  menucolor=blue,pagecolor=blue,%
  urlcolor=blue]{hyperref}
\else
\DeclareGraphicsExtensions{.eps, .jpg}
\usepackage[%
  breaklinks=true,%
  colorlinks=true,%
  linkcolor=blue,anchorcolor=blue,%
  citecolor=blue,filecolor=blue,%
  menucolor=blue,pagecolor=blue,%
  urlcolor=blue]{hyperref}
\fi

\begin{document}

\hugehead

\title{
  Bose--Einstein correlations in hadron-pairs from lepto-production on nuclei ranging from hydrogen to xenon
}

\author{ 
The HERMES Collaboration \medskip \\
A.~Airapetian$^{13,16}$,
N.~Akopov$^{27}$,
Z.~Akopov$^{6}$,
E.C.~Aschenauer$^{7}$,
W.~Augustyniak$^{26}$,
R.~Avakian$^{27}$,
A.~Avetissian$^{27}$,
E.~Avetisyan$^{6}$,
S.~Belostotski$^{19}$,
N.~Bianchi$^{11}$
% ...
}

\institute{ 
$^1$Physics Division, Argonne National Laboratory, Argonne, Illinois 60439-4843, USA\\
$^2$Istituto Nazionale di Fisica Nucleare, Sezione di Bari, 70124 Bari, Italy\\
$^3$School of Physics, Peking University, Beijing 100871, China\\
$^4$Department of Theoretical Physics, University of the Basque Country UPV/EHU, 48080 Bilbao, Spain and IKERBASQUE, Basque Foundation for Science, 48013 Bilbao, Spain\\
$^5$Nuclear Physics Laboratory, University of Colorado, Boulder, Colorado 80309-0390, USA\\
$^6$DESY, 22603 Hamburg, Germany\\
$^7$DESY, 15738 Zeuthen, Germany\\
$^8$Joint Institute for Nuclear Research, 141980 Dubna, Russia\\
$^9$Physikalisches Institut, Universit\"at Erlangen-N\"urnberg, 91058 Erlangen, Germany\\
$^{10}$Istituto Nazionale di Fisica Nucleare, Sezione di Ferrara and Dipartimento di Fisica e Scienze della Terra, Universit\`a di Ferrara, 44122 Ferrara, Italy\\
$^{11}$Istituto Nazionale di Fisica Nucleare, Laboratori Nazionali di Frascati, 00044 Frascati, Italy\\
$^{12}$Department of Physics and Astronomy, Ghent University, 9000 Gent, Belgium\\
$^{13}$II. Physikalisches Institut, Justus-Liebig Universit\"at Gie{\ss}en, 35392 Gie{\ss}en, Germany\\
$^{14}$SUPA, School of Physics and Astronomy, University of Glasgow, Glasgow G12 8QQ, United Kingdom\\
$^{15}$Department of Physics, University of Illinois, Urbana, Illinois 61801-3080, USA\\
$^{16}$Randall Laboratory of Physics, University of Michigan, Ann Arbor, Michigan 48109-1040, USA \\
$^{17}$Lebedev Physical Institute, 117924 Moscow, Russia\\
$^{18}$National Institute for Subatomic Physics (Nikhef), 1009 DB Amsterdam, The Netherlands\\
$^{19}$B.P. Konstantinov Petersburg Nuclear Physics Institute, Gatchina, 188300 Leningrad Region, Russia\\
$^{20}$Institute for High Energy Physics, Protvino, 142281 Moscow Region, Russia\\
$^{21}$Institut f\"ur Theoretische Physik, Universit\"at Regensburg, 93040 Regensburg, Germany\\
$^{22}$Istituto Nazionale di Fisica Nucleare, Sezione di Roma, Gruppo Collegato Sanit\`a and Istituto Superiore di Sanit\`a, 00161 Roma, Italy\\
$^{23}$TRIUMF, Vancouver, British Columbia V6T 2A3, Canada\\
$^{24}$Department of Physics, Tokyo Institute of Technology, Tokyo 152, Japan\\
$^{25}$Department of Physics and Astronomy, VU University, 1081 HV Amsterdam, The Netherlands\\
$^{26}$National Centre for Nuclear Research, 00-689 Warsaw, Poland\\
$^{27}$Yerevan Physics Institute, 375036 Yerevan, Armenia\\
} 

\date{DESY Report 15-074 / Compiled: \today\  / Version: 5.0}

\titlerunning{Paper Tag: BEC}
\authorrunning{The HERMES Collaboration}

\abstract{
Bose--Einstein correlations of like-sign charged hadrons
produced in deep-inelastic electron and positron scattering 
are studied
in the HERMES experiment using nuclear targets of $^1$H, $^2$H, $^3$He,
$^4$He, N, Ne, Kr, and Xe.
A Gaussian approach
is used to parametrize a two-particle correlation function determined
from events with at least two charged hadrons of the same sign charge. This correlation function is
compared to two different empirical distributions that do not include the Bose--Einstein
correlations.
One distribution is derived from unlike-sign hadron pairs, and the second is
derived from mixing like-sign pairs from different events. 
The extraction procedure used 
simulations incorporating the experimental setup
in order to correct the results for spectrometer acceptance effects, and
was tested using the distribution of unlike-sign hadron pairs. 
Clear signals of Bose--Einstein correlations for all target nuclei without
a significant variation with the nuclear target mass are found. 
Also, no evidence for
a dependence on the invariant mass $W$ of the photon-nucleon system is found when 
the results are compared to those of previous experiments. 
}
\maketitle

\subsection*{Introduction}
Hadron production in deep-inelastic scattering (DIS) of leptons off nuclei
is a powerful tool to study the quark hadronization process. 
The distance scale over which a struck quark that received a sufficiently large
energy-momentum transfer from an incident lepton develops into a colorless
hadronic particle extends well beyond the size of a single nucleon.
Therefore, the distribution of hadrons in the final state may be modified by interactions of 
the developing hadronic state with the nuclear medium outside the struck nucleon.
In general, this intermediate state is some mixture of quarks and 
gluonic fields that have not reached their asymptotic (confined) states, and so any
modification should depend on the evolution of that state.
Similarly the fully formed hadron may still pass through the nuclear medium and be subject to rescattering
processes (see, e.g., Ref.~\cite{hadrnucl}).  

One means of studying the final hadronic state is the use of Bose--Einstein 
correlations (BEC) in the distribution of bosons, in particular pions. These
correlations arise from interference between different parts of the symmetrized wave function of identical bosons 
from incoherent sources.
This well-known technique of {\it intensity} interferometry
was first developed by Hanbury Brown and Twiss to measure stellar radii \cite{hbt}.
Its first use in particle physics, half a century ago, was
to study the $p\bar p$ annihilation process \cite{Ggoldobs,Ggold} with incident 
anti-protons of 1~GeV momentum. Since then many 
measurements of BEC have been performed in
hadron-hadron scattering experiments. In addition, several studies of BEC
in the $e^{+}e^{-}$ annihilation process have been performed (see, e.g., 
Ref.~\cite{alexander}), especially by the LEP
experiments. Measurements of BEC from deep-inelastic lepton scattering experiments are less
abundant. The results from experiments using charged 
leptons as incident particles can be found in Refs.~\cite{EMC,E665,H1,ZEUS,ZEUSkaons},
while the results from neutrino experiments are found in Refs.~\cite{SKAT,BBCN,NOMAD}.
Several reviews~\cite{alexander,weiner,kittel,zalewski} summarize the present 
theoretical and experimental knowledge of BEC. 
The theory of BEC in particle physics was originally developed in the papers
of Kopylov and Podgoretskii~\cite{kopylpod1,kopylov,kopylpod2} and 
Cocconi~\cite{cocconi}. 
It should be noted that most of the theoretical work has
focused on the understanding of BEC in heavy-ion collisions, in which a ``fireball''
source distribution,
created by the collision roughly at rest involving many parton elementary interactions, decays into hadrons. 
Only a few references consider the quite different case of fragmentation in DIS and $e^+e^-$ processes, 
in which quite different
hadron-momentum and spatial-source distribution might be assumed (see, e.g., Ref.~\cite{BjorkenN,ANDE}). 
Estimates of BEC  in $e^{+}e^{-}$ annihilation from 
string-fragmentation models~\cite{ANDE} indicate that correlation
parameters are mostly dependent on string-breaking parameters, because the 
strongest correlations are from pions resulting from adjacent breaks along
a string.

\begin{figure}[t]
\centering
\includegraphics[width=6.5cm]{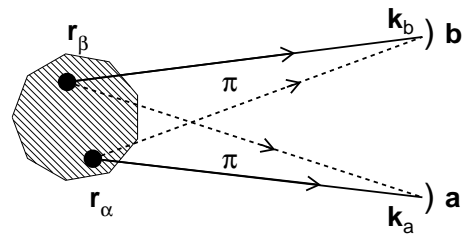} 
\caption{Schematic illustration of the Bose--Einstein effect.}
\label{fig:bepict}
\end{figure}

To better understand the underlying physics of BEC one may consider a simple example
of the emission and detection of two identical bosons, e.g., 
two pions, from points  $\vec{r}_\alpha$ and $\vec{r}_\beta$, 
which are observed with momenta $\vec{k_a}$ and $\vec{k_b}$ at detectors {\it a} 
and {\it b} (Fig.~\ref{fig:bepict}).
The two pions are indistinguishable and the total wave function of the two-pion
system must be symmetric under the exchange of them:
\beq
\label{eq:wavefunc}
\Psi_{2\pi} = \frac{1}{\sqrt{2}} \biggl( \Psi_{a\alpha}  \Psi_{b\beta} +  \Psi_{b\alpha}  \Psi_{a\beta} \biggr),
\eeq
where $\Psi_{a\alpha}$ is the wave function of a pion produced at point $\vec{r_{\alpha}}$ and 
observed at detector $\vec{a}$ while $\Psi_{b\beta}$ is the wave function of a pion produced at
point $\vec{r_{\beta}}$ and observed at detector $\vec{b}$. Assuming plane waves, i.e.,
$\Psi_{a\alpha} \approx \exp(i\vec{k_{a}}\vec{r_{\alpha}})$, one may obtain 
$ |\Psi_{2\pi}|^2 = 1 + \cos(\vec{\delta k}\cdot\vec{\delta r})$ with $\vec{\delta k}=\vec{k_{a}}-\vec{k_{b}} $
and $\vec{\delta r}=\vec{r_{\alpha}}-\vec{r_{\beta}} $. Thus the correlation function resulting from the 
interference
of the two terms in Eq.~\eqref{eq:wavefunc} will take the following form:
\beq
\label{eq:correlfunc}
R(\vec{k_{a}},\vec{k_{b}}) \propto 1 + \cos(\vec{\delta k}\cdot\vec{\delta r}).
\eeq
This expression shows that the BEC effect measures the projection of the spacial distance ($\vec{\delta r}$) between two
particle sources on the direction of the momentum difference ($\vec{\delta k}$) between the observed pions.
One can generalize two-point sources to a continuous space-time distribution of sources.
Experimentally this is achieved by finding the boson correlation function.
In BEC this correlation observable is defined in terms of the two-particle correlation function
\beq
\label{eq:defbec}
R(p_1, p_2) = D( p_1, p_2) / [D(p_1)\cdot D(p_2)] \, ,
\eeq
where $p_1$ and $p_2$ are the particle four-momenta, $D( p_1, p_2)$ 
is the two-particle probability density and $D(p_1)$, $D(p_2)$ are one-particle 
probability densities.

Typical analyses use models for the correlation function $R$ with a limited set of parameters.
In the case of fireball decay (heavy-ion collision), the
correlation function might be described by a (spherical) 
Gaussian distribution in space with an independent parametrized exponential decay
in time. However, such an approach is not Lorentz invariant.
Generally the Goldhaber parametrization~\cite{Ggold} is a more convenient approach, 
in which the distribution is
a function of the Lorentz-invariant quantity $T^2$ of the hadron pair,
where ${T}^2 = -(p_1-p_2)^2 =  S -  4m_\pi^2 $ with $S$ the squared invariant mass of 
the pair and $m_\pi$ the pion mass.
In this parametrization the correlation function $R$ takes
the form
\beq
\label{eq:goldhaber}
 R(T) = 1 + \lambda \cdot e^{-T^2 r_G^2}.
\eeq
This parametrization corresponds to a Gaussian shape 
of the particle source distribution of size $r_G$ in the center of mass of the pair.
The additional parameter $\lambda$, the chaoticity (or incoherence), has been introduced 
to account for a possible incoherent contribution of the pion emitters.
Completely coherent pion sources lead to the absence of correlations among the bosons 
($\lambda = 0$)~\cite{gamow,Bowler}, while in the simplest theoretical treatment, completely incoherent sources
lead to $\lambda = 1$. Subsequent theoretical work has shown that a number
of different effects can modify the value of $\lambda$ in both directions, greater and smaller than 
unity (see, e.g. Ref.~\cite{weiner}). 
Examples of such effects are 
the presence of decay products of long-lived resonances in addition to directly produced bosons in the final state,
final-state interactions (FSI), and the deviation of the exact shape of the source distribution from the assumed Gaussian form.
Other experimental effects can also influence the experimentally measured value of $\lambda$, 
often due to the purity of the boson sample as determined by the
quality of the particle identification within the experiment. For these reasons, it is
difficult to compare the results of $\lambda$ between different experiments, even at 
identical kinematic conditions.

Note that the Goldhaber parametrization does not correspond to
any well developed theoretical approach but rather serves as a convenient tool for the 
comparison of experimental results. 
The parametrization describes the shape of the
distribution in $T$ satisfactorily.
However, depending on the physics of the reaction, a 
rigorous treatment of the coherent and 
incoherent sources requires modification of the functional form of the correlation
function (see, e.g., Ref.~\cite{GEHW}).

Given these remarks, one must note that
there are additional points to consider in a full treatment of BEC.
Initial correlations among the bosons are affected by FSI
between the produced particles as well as with the production environment, both
via the Coulomb interaction and hadronic interactions.
The long-range Coulomb FSI is quite often accounted for by introducing 
a multiplicative Gamow factor \cite{gamow}.
It increases (decreases) the two-particle density for opposite-(like-)sign particles. 
The correction factor is essential only at very small values of $T$ and very quickly 
approaches unity with increasing $T$. 
For pion pairs at $T = 0.05$~GeV the correction factor differs from unity by only 
about $1.5\%$. There are other more elaborate calculations that 
predict an even smaller magnitude for the required correction and also 
examine its model dependence.
Short-range strong FSI between the two identical pions may also influence
their correlation (see, e.g., Ref.~\cite{alexander,weiner,kittel,zalewski,Bowler1}); 
without clear theoretical
estimates for the kinematics of the present experiment 
it was chosen not to attempt any correction for both long-range and short-range FSI.

Effects that can alter the pair correlation within the nuclear environment are the focus of
this study of BEC for nuclear targets ranging from hydrogen
to xenon. Measurements with the same experimental apparatus and kinematics
help to minimize possible systematic bias in the observed target-mass dependence.
In DIS a difference in the size of the particle emission region 
could exist for pions produced off a free nucleon as compared to that off 
bound nucleons. In addition, interactions of the struck bare quark, during fragmentation, or of the fully
developed hadron with the nuclear environment could alter both the apparent
size of the emission region and the amount of incoherence.
For example, a common simple assumption is that the correlations of a pair of 
identical pions are determined by the relative positions of their last 
scattering points, which thereby play the role of independent particle sources. If the
pions scatter from the nuclear matter, one would expect an increase
in the size of the emission source as a function of target radius. Another example
would be the increasing probability of gluon radiation within the nucleus leading to a
change in the pion-pair correlations relative to hadronization in free space.

The influence of nuclear re-scattering processes on BECs in heavy-ion collisions 
was studied in Ref.~\cite{kapusta} and effects on the source size of 15-20\% were found.
No estimates exist for the case of lepton-nucleus hadron production.
Earlier experimental studies of BECs from DIS by nuclei are quite limited.
The BBCN Collaboration \cite{BBCN} found the BEC parameters $r_G$
and $\lambda$ to be independent of the atomic mass. 
These measurements, however, are limited to three light nuclei, 
${}^1$H, ${}^2$H, and Ne. 

There is significant evidence that the nuclear medium affects
the hadronization processes 
in lepton scattering at HERMES energies. Results on this subject have been presented in a number of papers
~\cite{hadrnucl,coherence,hadronform,quarkfrag,piKfrag,doublehad,ptbroad,PhysRevD.90.072007}.
In particular,
the transparency of the $^2$H, $^3$He, and $^{14}$N nuclei to exclusive incoherent 
$\rho^0$ electro-production has been measured \cite{coherence}
and significant dependences on the coherence length were found.
A series of papers
~\cite{hadrnucl,hadronform,quarkfrag,piKfrag} is devoted to the investigation of the hadron
multiplicity variation in different kinematic regions and its dependence on the 
target atomic mass $A$ up to xenon.
The most prominent features of the data are an increased hadron attenuation with 
increasing value of the mass number $A$ of the nucleus
and the attenuation becoming smaller (larger) with increasing values of $\nu$~$(z)$,
where $\nu$ is the energy of the virtual photon in the laboratory system, 
and $z=E_h/\nu$ is the fractional hadron energy.

The influence of the nuclear medium on the ratio of double-hadron to single-hadron
yields in DIS was also investigated \cite{doublehad}. Nuclear effects are clearly observed but
with substantially smaller magnitude as well as reduced $A$ dependence compared to the 
single-hadron multiplicity ratios.
The first detailed study of the dependence on the target nuclear mass of the
average squared transverse momentum $\langle p_t^2 \rangle$ of hadrons produced in 
deep-inelastic lepton scattering is described in \cite{ptbroad}. It is found that 
the average squared transverse momentum is increasing with the atomic mass number.

In short, several studies with different nuclear targets performed at HERMES show
significant modifications of 
the hadron observables within the nuclear medium as compared to the 
results on a proton/deuteron target.

\subsection*{Experiment}
The present measurement of the BEC was performed with the HERMES
spectrometer~\cite{hermesdetector} using the 27.6~GeV polarized lepton (electron/positron)
beam stored in
the HERA ring at DESY. The spectrometer consisted of two identical halves
above and below the lepton beam line. 
The scattered lepton and the
produced hadrons were detected within an angular acceptance of
$\pm$170~mrad horizontally, and $\pm$(40--140)~mrad vertically.

All the targets were internal to the lepton
storage ring and consisted of polarized or unpolarized ${}^1$H, ${}^2$H, and ${}^3$He, 
or unpolarized ${}^4$He, N, Ne, 
Kr, or Xe gas injected into a thin-walled open-ended
tubular storage cell. Target areal densities up to 1.4 $\times$ 10$^{16}$
nucleons/cm$^2$ were obtained for unpolarized gas corresponding to
luminosities up to 3 $\times$ 10$^{33}$ cm$^{-2}$ s$^{-1}$.
The luminosity was measured using elastic scattering of the beam leptons off the electrons
in the target gas, Bhabha scattering for a positron beam and 
M\o ller scattering for an electron beam \cite{lumi}.

The trigger was formed by a coincidence between the signals from three
scintillator hodoscope planes, and a lead-glass calorimeter where a minimum
energy deposit of 3.5~GeV (1.4~GeV) for unpolarized (polarized)
target was required. The scattered leptons were identified using
a transition-radiation detector, a scintillator pre-shower counter, an
electromagnetic calorimeter, and a threshold gas Cherenkov counter.
In 1998 the threshold Cherenkov counter was replaced by a ring-imaging
Cherenkov detector (RICH).

\subsection*{Analysis}
Scattered leptons are selected by imposing constraints on the squared
four-momentum of the virtual photon, $Q^2>1$~GeV$^2$, and on the invariant
mass of the photon-nucleon system, $W^2>$ $10$~GeV$^2$.
The constraint on $W^{2}$ is applied in order to ensure the predominance of 
multiple particle production in the DIS events.

The results of this study are
based on data collected by the HERMES Collaboration between the years 1996 and 2006.
The yields from polarized targets are summed
over both target spin orientations. The yields from all targets 
are summed over both (longitudinal) beam polarization states.

Events with only one identified lepton of the same charge as the beam lepton 
and momentum larger than $3.5$~GeV are accepted. The presence of at least two  
charged hadrons  with  momenta $2.0$~GeV $<$ $ p _ {h} < 15 $~GeV
is required for further analysis.
The total numbers of such multi-hadron DIS events, $N_{ev}$, the numbers of like-sign, $N^{like}$, 
and unlike-sign,
$N^{unlike}$, hadron pairs 
available for the analysis are given in Table~\ref{tab:events} for each target.

\begin{table}[t]
\centering
\caption{Number of DIS events with more than one detected hadron, $N_{ev}$, 
the number of like-sign hadron pairs, $N^{like}$, and of 
unlike-sign hadron pairs, $N^{unlike}$, that meet the kinematic requirements for each target.}
\label{tab:events} 
\begin{tabular}{|c|r|r|r|}
\hline
& & &   \\
 ~Nucleus~ &$ ~~~~~N_{ev}$~~~~ & ~~~~$N^{like}$~~~ & ~~~$N^{unlike}~~ $\\
& & &  \\
\hline
& & &  \\
${}^1$H &~1145046 &~~478946 &958185  \\
${}^2$H &1297356 &680143 &~1178797  \\
${}^3$He &34391 &15295 &29165 \\
${}^4$He &79776 &30539 &59244 \\
N &92968 &41112 &78402 \\
Ne &175594 &75898 &146145 \\
Kr &{211456} &91391 &172946 \\
Xe &106274 &46130 &87125 \\
&&& \\
\hline
\end{tabular}
\end{table}

In this analysis all charged hadrons are considered to be pions.
Simulations using the PYTHIA~6.2 event generator~\cite{PYTHIA}, tuned to provide an accurate 
description~\cite{leibing} of semi-inclusive deep-inelastic hadron lepto-production in the HERMES 
kinematic region, show that the observed charged hadrons are distributed
in relative proportion $\pi / K / p(\bar{p}) = 78\% / 12\% / 10\%$.
In the same simulations one finds that $55\%$ of like-sign hadron pairs and $66\%$ of 
unlike-sign hadron pairs are truly pion pairs. This results in a ``dilution'' of the 
parameter $\lambda$ under the assumption that the non-pion pairs do not contribute to the BEC.
The value of $\lambda$ in this analysis is expected to be smaller than about 
$0.5$. Kaon like-sign pairs contribute only
$2\%$ while their unlike-sign pairs contribute about $4\%$.

Experimentally, it is difficult to measure the inclusive single particle 
spectrum required to determine the probability density $D(p)$ for all possible momenta $p$ in the formal
definition of the correlation function $R(p_1,p_2)$.
A common practice is to substitute 
the two one-particle probability distributions $D(p_1)\cdot D(p_2)$ with a two-particle probability density 
reference distribution $D_{r}(p_1, p_2)$.
This reference distribution is constructed from experimental two-particle distributions that 
do not have any BECs. The experimental correlation function is then defined as 
\beq
\label{eq:corrfunc}
R(p_1, p_2) = D( p_1, p_2) / D_{r}(p_1, p_2) .
\eeq
The Goldhaber parametrization usually used includes an additional normalization
parameter $\gamma$ and a polynomial function ${\cal P}(T)$ to describe the long-range correlations at large $T$
and has the form
\beq
\label{eq:goldexp}
R (T) = \gamma \cdot [ 1 + \lambda \cdot e^{- T^2 r_G^2} ] \cdot {\cal P}( T ).
\eeq 
The long-range correlations at large $T$ may appear 
due to charge and energy conservation, phase-space constraints and 
imperfections in the reference sample. 
The form of the polynomial ${\cal P} ( T )$ used to model these long-range correlation
effects is taken to be $(1 + \delta \cdot T^2)$.
A linear dependence $(1 + \delta^\prime \cdot T)$ is used to estimate the influence of
the chosen form on the final results. The parameter $\delta$ and $\delta^\prime$ are free parameters
like $r_G$, $\lambda$, and $\gamma$.

The magnitude of the two-particle BEC is measured
by comparison of the experimental distribution with a reference sample distribution 
[see Eq.~\eqref{eq:corrfunc}].
The method of constructing the reference sample is the main source of systematic uncertainty,
especially since the multiplicity of hadrons in the HERMES experiment
is relatively low. 
One of the main problems in measuring BEC is 
the evaluation of biases caused by an imperfect reference sample,
thus it is desirable to use at least two different approaches to construct a reference sample
to cross-check the correlation results.

Two of the most widely used methods to construct a reference sample are employed here:
\begin{itemize}
\item Method of event mixing ($MEM$), 
\item Method of unlike-sign pairs ($MUS$). 
\end{itemize}
In $MEM$ hadron-pair distributions of the same charge are created by using hadrons
from different events, while in $MUS$ hadrons with different charge from the same
event are used.
Other methods can be found in the literature, each with its own shortcomings.
The main technical difficulty of the two methods chosen here
is the violation of momentum and energy conservation in the kinematic event topology when selecting two 
hadrons from different events in the case of $MEM$,
and the contribution of events from resonances that are not 
present in the like-sign distribution in the case of $MUS$.
The systematic effects associated with these two reference samples
are studied using the PYTHIA-based Monte Carlo simulation of the HERMES experiment discussed above, 
the inclusion of quantum interference of BEC not being enabled in the PYTHIA event generation.

The construction of the reference sample using the $MUS$ is 
done by forming the distribution in $T$ of 
all unlike-sign hadron pairs, requiring the 
same constraints as for the like-sign pairs. For the $MEM$,
to construct a sample of uncorrelated hadron pairs, a 
combination of charged hadrons from two different DIS events is used.
The first hadron of a pair is taken from one event 
while the second hadron is taken from another event.
Care must be taken to conserve collinearity of the virtual-photon vectors
$\vec{q}_1$ and $\vec{q}_2$ from these two different events.
For each event, the momentum vector of the total hadronic system must lie in the direction
of the virtual photon  $\vec {q} = \vec{p_e} - \vec{p}_{e}^{\prime}$,
where $\vec{p_e}$ and $\vec{p}_{e}^{\prime}$ are the momenta of the incident and 
scattered leptons.
To conserve collinearity
the momenta of all hadrons in the second event must be rotated in such a way that the total 
hadronic momentum is aligned along the direction of $\vec{q}_1$ of the first event.
\begin{figure}[t]
\centering
\includegraphics[width=8.5cm]{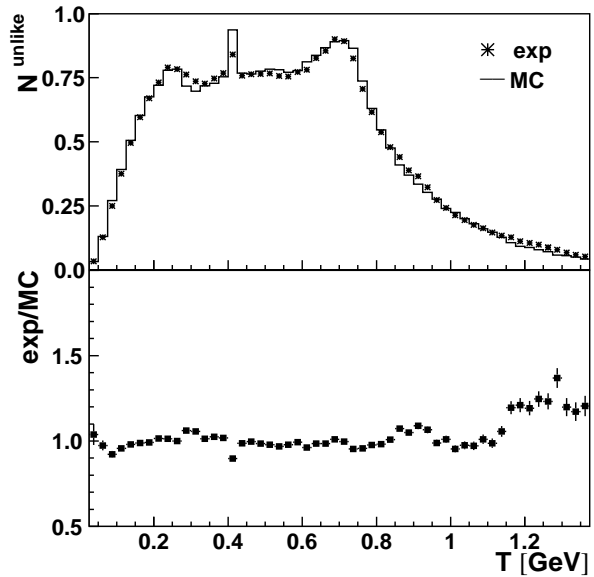}
\caption{Top panel: normalized experimental (stars) and 
simulated (line) distributions for unlike-sign hadron pairs 
as a function of the variable $T$. 
Bottom panel: the ratio of these experimental (exp) and simulated (MC) distributions.}   
\label{fig:mcdatacomp} 
\end{figure}

The quality of the HERMES simulation with respect to the description of the measured
unlike-sign hadron pair
distribution is demonstrated in Fig.~\ref{fig:mcdatacomp} for
a sample of hydrogen target data. The top panel shows the 
experimental $T$ distribution of unlike-sign hadron pairs (exp) in
comparison with the simulated data for $h^+h^-$ pairs (MC).
The bottom panel shows their ratio.
The figure demonstrates good agreement between the experimental and
simulated distributions. Based on this agreement, the simulation results are used
to further 
reduce the systematic biases of the reference sample and experimental distributions through the use of a 
double-ratio definition for the correlation function $R(T)$. For the two methods in this analysis,
\begin{equation}
\begin{aligned}
&R^{MEM} &= (like/mixed)^{exp}  & / \, (like/mixed)^{MC},\\
&R^{MUS} &= (like/unlike)^{exp} & /  \, (like/unlike)^{MC}.\\
\end{aligned}
\end{equation}
Dividing the experimental ratios by the simulated ratios is expected to reduce biases 
resulting from the violation of kinematic constraints in the $MEM$ and from resonance contamination 
in the $MUS$, since these biases also exist in the simulated event distributions.

\begin{figure}[tb]
\centering
\includegraphics[width=7.0cm]{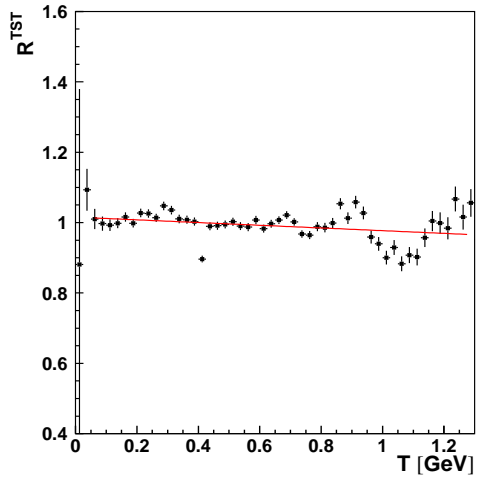}
\caption{Consistency check of the two chosen reference samples. The quantity $R^{TST}$ is defined in the text.
The curve is a linear fit to the data for $T$ between 0.05~GeV and 1.3~GeV.}
\label{fig:unlike-test}
\end{figure}

As a test of the validity of $R(T)$ using the simulation results with a
double ratio, the $MEM$ was used with unlike-sign hadron pairs from the
hydrogen data sample to construct the double ratio
\begin{equation}
R^{TST} = (unlike/mixed)^{exp} / \; (unlike/mixed)^{MC}, 
\end{equation}
shown as a function of $T$ in
Fig.~\ref{fig:unlike-test}. This test ratio is expected to have
no BECs, and ideally would 
have a value of unity over the entire $T$ range. At very low $T$ ($\leq 0.05$~GeV), at 
$T \approx 0.4$~GeV, and at $T > 0.9$~GeV this double
ratio deviates from unity significantly. As shown by a linear fit to $R^{TST}$ there is a slight 
linear dependence over most of the range of $T$, indicating some small residual bias.
The deviation near 0.4~GeV in the simulation is likely due to insufficient description of $K_S$ production, 
which contributes to the $N^{unlike}$ distributions (see Fig.~\ref{fig:mcdatacomp}). 
The deviations at very low and at large $T$ likely arise from
some combination of effects in both the simulation of the $MUS$ and the $MEM$
construction of the reference sample. The very low $T$ region, $T < 0.05$~GeV, 
of the double ratio distributions is excluded from further analysis due to lack of statistics.
A fit to the correlation function $R^{TST}$ (shown in Fig. \ref{fig:unlike-test}) with
the Goldhaber parametrization [Eq.~\eqref{eq:goldexp}]
over the range $0.05$~GeV$<T<1.30$~GeV gives $\lambda$=0.000$\pm$0.003 and $r_G=$0.0$\pm$1.4~fm, suggesting
that the fluctuations at large $T$ and the slight non-zero linear dependence on $T$ do not cause a 
significant
bias of the extracted parameters $\lambda$ and $r_G$.

\subsection*{Results}
\begin{figure}[t]
\centering
\includegraphics[width=7.0cm]{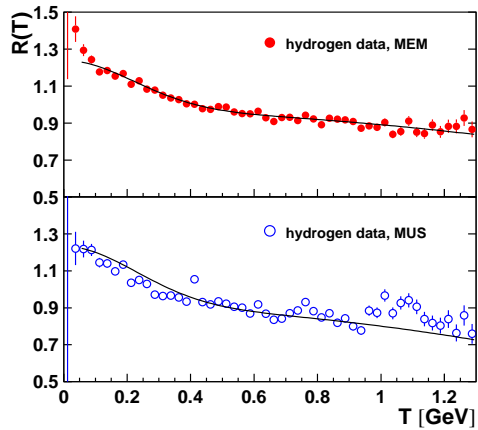}
\caption{Double ratio correlation function for like-sign hadron pairs
obtained with $MEM$ and $MUS$ based on hydrogen target data.}
\label{fig:corrfunct}
\end{figure}

The double-ratio correlation functions obtained from hydrogen data are shown
in Fig.~\ref{fig:corrfunct} for both types of the reference sample.
The curves in the figure are results of fits using the Goldhaber
parametrization [Eq.~\eqref{eq:goldexp}]. The fits are performed over the range of
0.05~GeV $<T<$ 1.30~GeV. 
The values for the two parameters obtained from the fits are given in Table~\ref{tab:hy_results}.

\begin{table}[b]
\centering
\caption{Results for the Goldhaber  parametrization fitted to the HERMES hydrogen data, both for the mixed-event method ($MEM$) and the method of unlike-sign pairs ($MUS$).}
\label{tab:hy_results}
  \begin{tabular}{|c|c|}
    \hline
    & \\
  Method &  Goldhaber parameters\\
  \hline
  & \\
  $MEM$ & $r_G=0.64\pm 0.03$(stat)${}^{+0.04}_{-0.04}$(sys) fm \\
        & $\lambda=0.28\pm 0.01$(stat)${}^{+0.00}_{-0.05}$(sys)$\mbox{ }\mbox{ }\mbox{ }$\\
  & \\
  \hline
  & \\
  $MUS$ & $r_G=0.72\pm 0.04$(stat)${}^{+0.09}_{-0.09}$(sys) fm \\
  & $\lambda=0.28\pm 0.02$(stat)${}^{+0.02}_{-0.04}$(sys)$\mbox{ }\mbox{ }\mbox{ }$\\
  & \\
  \hline
\end{tabular}
\end{table}
The systematic uncertainties are estimated by variations of the
fit range in $T$, the bin width, and the 
polynomial form for the long-range correlations term, i.e., 
using a linear dependence $(1+\delta^\prime T)$.
The results of the two different methods are 
consistent (see Table~\ref{tab:hy_results}). Values of the fit parameter $\delta$ from the quadratic form of ${\cal P}( T )$
are $-0.08 \pm 0.01$ and $-0.05 \pm0.01$ respectively for the $MEM$ and $MUS$.

\begin{figure}[t]
\centering
\includegraphics[width=8.0cm]{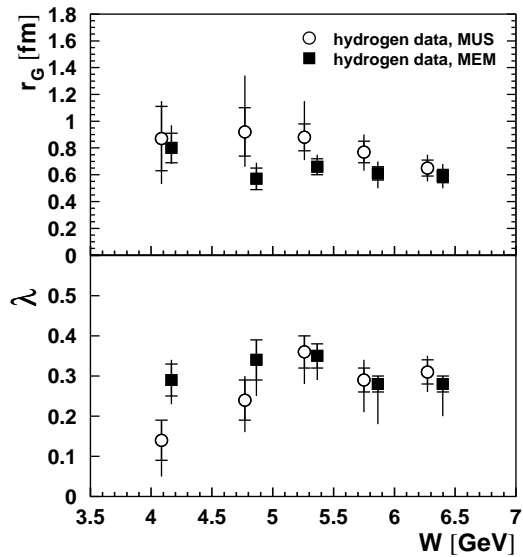}
\caption{Parameter $r_G$ (top panel) and  $\lambda$ (bottom panel) as a function of $W$,
obtained with $MEM$ and $MUS$ methods on hydrogen. The inner and outer error bars indicate
the statistical and total uncertainties. For the latter the statistical and systematic 
uncertainties are added in quadrature.}
\label{fig:w_rl}
\end{figure}

The kinematic dependence of the BEC parameters on the invariant mass $W$ of the photon-nucleon
system has been 
studied for the hydrogen target data sample.
In Fig.~\ref{fig:w_rl} the resulting parameters $r_G$ and $\lambda$ are presented 
for like-sign hadron pairs as a function of $W$ 
obtained with the $MEM$ and $MUS$ methods.
Within the present systematic and statistical uncertainties there is no clear
dependence of the parameters on the invariant mass $W$ in this range. Previous measurements
from the HERA H1 experiment~\cite{H1} over a broad range at high $W$ (65~GeV $<$ $W$ $<$ 240~GeV)
found only slight evidence of an increase in $r_G$. 

\begin{figure}[t]
\centering
\includegraphics[width=8.0cm]{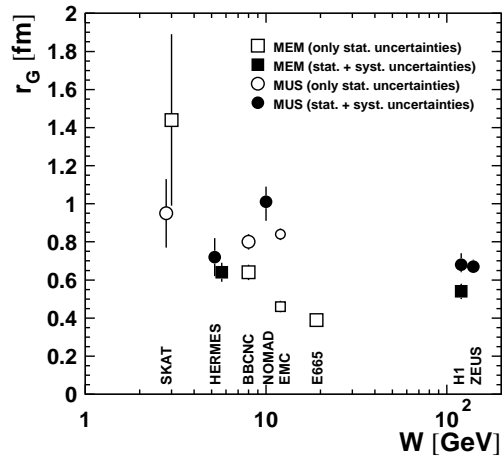}
\caption{Goldhaber radius $r_G$, as a function of $W$, obtained in 
lepton nucleon scattering experiments~\cite{EMC,E665,H1,ZEUS,SKAT,BBCN,NOMAD}. Different markers are 
used to indicate the different methods for the construction of the reference sample
and the kinds of uncertainties included.}
\label{fig:other_W_r}
\end{figure}

As mentioned above, BEC has been studied in a number of
lepton-hadron and $e^{+} e^{-}$ experiments.
The Goldhaber parametrization is used in most of these analyses. 
The parameter $r_G$ as a function of the average value of $W$ in lepton-nucleon scattering
experiments 
is shown in Fig.~\ref{fig:other_W_r}. The parameter $\lambda$ for a given experiment may 
depend on the hadron fractions and on the experimental details, hence
the results of $\lambda$ obtained here are not compared to those in other measurements. 
In the majority of these experiments
the extracted values of $r_G$
depend upon the method of the construction of the reference sample. 
Even for a single experiment, e.g., EMC,
the parameter $r_G$ obtained with the $MUS$ is twice as large as
that obtained with $MEM$.
From Fig.~\ref{fig:other_W_r} no clear dependence of the parameter $r_G$
on $W$ can be deduced, from neither methods (MEM and MUS).
The following conclusions are drawn from a comparison of these results from
the different experiments:
\begin{enumerate}
\item Most values of the parameter $r_{G}$ are in the range of 
      0.4~fm to 1.0~fm.
\item The results strongly depend on the choice of the reference sample.
   Analyses of the same data set with different reference samples often give 
incompatible results for $r_{G}$ (and $\lambda$).
\item The $MUS$ typically gives higher values for the parameter $r_{G}$ than the $MEM$.
\end{enumerate}
The HERMES results on hydrogen are in
general agreement with those 
of previous lepton-nucleon scattering experiments over a broad range in $W$, and agree well with the
BBCN neutrino experiment, which is at a slightly higher mean $W$ than HERMES. 
Similar results are seen in $e^+e^-$ collisions at LEP (see Ref.~\cite{alexander}).

\begin{figure}
\centering
\includegraphics[width=8.0cm]{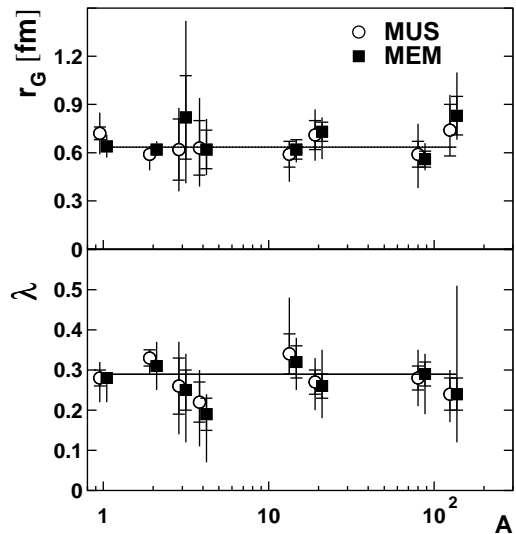}
\caption{The parameters $r_G$ (top panel) and $\lambda$ (bottom panel) 
are shown as a function of the target atomic mass $A$. 
The inner part of the error bars indicate the statistical uncertainty and the
total error bars have systematic uncertainties added in quadrature. 
The horizontal lines correspond to the average value of the parameters.}
\label{fig:R_A} 
\end{figure}

A possible nuclear dependence in BEC was examined using an extensive HERMES data set (cf.~Table~\ref{tab:events}).
The correlation function for like-sign hadron pairs produced in 
scattering off the nuclear targets $^2$H, $^3$He, $^4$He, N, Ne, Kr, and Xe was determined
using the same approximate parametrization as given in Eq.~\ref{eq:goldexp}. 
Systematic uncertainties are estimated separately for each target
and each reference sample ($MEM$ and $MUS$). 
The parameters $r_G$ and $\lambda$ are presented in Fig.~\ref{fig:R_A} as a 
function of the target atomic mass $A$. 
No dependence of these parameters on target atomic mass is observed within the estimated uncertainties.
Fit results with a constant over the whole range of the atomic mass for the 
four sets of data points are presented in Table~\ref{tab:A-dependence}.
\begin{table}[b]
\centering
\caption{Fit of a constant to the Goldhaber parameters as a function of the target atomic mass $A$. 
Results are given for both the mixed-event method ($MEM$) and the method of unlike-sign pairs ($MUS$).}
\label{tab:A-dependence}
\begin{tabular}{|c|c|c|}
\hline
& & \\
Method & Value & $\chi^2$/NDF \\
\hline
& &\\
$MEM$ &  $r_G = 0.634 \pm 0.017$~fm & 1.5 \\
      & $\lambda = 0.289\pm 0.006\mbox{ }\mbox{ }$  & 2.1 \\
& & \\
\hline
& &\\
$MUS$ & $r_G = 0.636 \pm 0.021$~fm & 1.2 \\
      & $\lambda =  0.289 \pm 0.011\mbox{ }\mbox{ }$ & 1.4 \\
& &\\
\hline
\end{tabular}
\end{table}
Here, the total uncertainty of each particular point is taken as 
the quadratic sum of statistical and systematic uncertainties.
The parameters extracted with the two reference samples are in good agreement.

To date there are no theoretical estimates for the magnitude of nuclear effects on BEC 
in DIS. 
In the absence of some hitherto unknown effect of multi-particle correlations, 
hadrons produced are expected to interact
with the nuclear medium.
Within the sensitivity of this experiment no clear dependence of the parameters
$\lambda$ and $r_G$ on the target atomic mass is observed, consistent with earlier results by the BBCN Collaboration~\cite{BBCN}. 
This is similar to the rather weak dependence of the double-hadron yields on the target atomic mass observed at HERMES~\cite{doublehad},
in contrast to much stronger effects observed in the distributions of single-hadron yields~\cite{hadrnucl,hadronform,quarkfrag,piKfrag}.

In conclusion, a study of the Bose--Einstein correlations between two like-sign hadrons
produced in semi-inclusive deep-inelastic electron/positron scattering off nuclear targets
ranging from hydrogen to xenon has been carried out.
Two different methods of constructing the reference sample are used in this study, and Bose--Einstein
correlations are clearly observed in all the data samples. 
The results obtained using the two reference sample methods are in good agreement, suggesting that 
most of the systematic
uncertainties connected with the construction of the reference samples are taken into account by the use of
double ratios corrected via an accurate experimental simulation.
Within the total experimental uncertainties,
no dependence of the parameters $r_G$ and $\lambda$ on the target 
atomic mass is observed.

\begin{acknowledgement}
We gratefully acknowledge the \desy\ management for its support and the staff
at \desy\ and the collaborating institutions for their significant effort.
This work was supported by 
the Ministry of Education and Science of Armenia;
the FWO-Flanders and IWT, Belgium;
the Natural Sciences and Engineering Research Council of Canada;
the National Natural Science Foundation of China;
the Alexander von Humboldt Stiftung,
the German Bundesministerium f\"ur Bildung und Forschung (BMBF), and
the Deu\-tsche Forschungsgemeinschaft (DFG);
the Italian Istituto Nazionale di Fisica Nucleare (INFN);
the MEXT, JSPS, and G-COE of Japan;
the Dutch Foundation for Fundamenteel Onderzoek der Materie (FOM);
the Russian Academy of Science and the Russian Federal Agency for 
Science and Innovations;
the Basque Foundation for Science (IKERBASQUE) and the UPV/EHU under program UFI 11/55;
the U.K.~Engineering and Physical Sciences Research Council, 
the Science and Technology Facilities Council,
and the Scottish Universities Physics Alliance;
as well as the U.S.~Department of Energy (DOE) and the National Science Foundation (NSF).
\end{acknowledgement}

\end{document}